\documentclass{PoS}

\title{WIMP theory review}

\ShortTitle{WIMP theory review}

\author{\speaker{FARINALDO queiroz}%
        \thanks{I am grateful to the organizers for the invitation. I acknowledge support from MEC and ICTP-SAIFR FAPESP grant 2016/01343-7.}\\
       Max Planck Institute for Nuclear Physics - Heidelberg, and\\
       International Institute of Physics, Federal University of Rio Grande do Norte,
Campus Universit\'ario, Lagoa Nova, Natal-RN 59078-970, Brazil\\
       E-mail: \email{queiroz@mpi-hd.mpg.de}}

\abstract{The complementarity of direct, indirect and collider searches for dark matter has improved our understanding concerning the properties of the dark matter particle. In this short review, we present a step toward the fundamental nature of dark matter with direct detection experiments only, go through some of the potential dark matter signals in gamma-rays, x-rays, and anti-matter, and lastly discuss the prospects of WIMPs in the next decade.}

\FullConference{The European Physical Society Conference on High Energy Physics\\
		5-12 July, 2017\\
		Venice}

\begin{document}

\section{Introduction}

The presence of dark matter has been ascertained through its gravitation effects by many observations \cite{Bertone:2016nfn}. Today, we have an accurate description of the universe content. Dark matter accounts for about 85\% of the matter content of the universe and roughly 27\% of the entire energy density. Unfortunately, we have not observed any robust signals of dark matter interactions with ordinary matter. Therefore, its nature constitutes one of the most exciting mysteries in science. For this reason, the literature is flooded with dark matter candidates with all sort of possible signatures. Here we will be concentrated on the most popular one, WIMPs (Weakly Interacting Massive Particles). 

Weak interactions refer to the interactions governed by the $SU(2)_L$ gauge bosons in the Standard Model, but we emphasize that despite WIMPs also refer to weak interactions, they are not necessarily particles governed by the $SU(2)_L$ gauge group. They are part of a much larger picture. This misinterpretation of what interactions WIMP can have might lead to wrong conclusions concerning the status of the WIMPs as dark matter candidates \cite{Arcadi:2017kky}. Anyway, WIMPs are particularly attractive because they are linked to physics not far from the electroweak scale, and their popularity is closely related to the paradigm of thermal decoupling. 

We have learned with the success of cosmology of the early universe that thermal decoupling is a powerful tool to explain observables in our universe. Indeed, the thermal decoupling furnishes detailed predictions for the abundances of light elements and the cosmic microwave background radiation. Hence, it is plausible to assume that the dark matter particle belonged to a thermal history. Under that assumption, an annihilation cross section around the weak scale arises rather naturally and independently of the quantum numbers of the dark matter particle. It is striking that I have informed neither the spin nor the underlying particle physics model to come to this conclusion.  Indeed, the dark matter relic density is naively expected to be,

\begin{equation}
\Omega_{DM} h^2 \sim \frac{\alpha^2_{em}}{M_W^2},
\label{DMomega}
\end{equation}where $\alpha_{em}$ is the electromagnetic fine structure constant, and $M_W$ the mass of the Standard Model (SM) gauge boson W. Since $\alpha^2/M_W^2$ is typically the weak-scale cross-section, then if the WIMP features a weak scale pair annihilation cross section it can naturally lead to the correct relic density. This fact is true if they belonged to a thermal history and their masses are at the weak scale. The expression above for the abundance is far from accurate because there are many important ingredients that we have been intentionally set aside in the discussion that significantly change the final dark matter abundance in Eq.\ref{DMomega} such as resonance effects, coannihilations, and kinematic threshold effects \cite{Griest:1990kh}. 

Another important aspect of WIMPs is that they were not postulated out of thin air simply to address the dark matter or to fit some data set, they actually arose as a natural consequence in models that address other important theoretical problems such as the hierarchy problem, differently than many other dark matter candidates in the literature. 

In summary, from the theoretical point of view WIMPS are desirable. On the experimental side, WIMPs are also very appealing. Using the weak scale as guiding principle one should expect a WIMP signal at current and planned experiments in most models. Though experiments are looking for the dark they more or less know where to find it. It is a falsifiable dark matter candidate and this is key for an experimentalist. If one starts searching for WIMPs outside the box of weak-scale interactions than one will be faced with the fact that WIMPs might be around the corner but may also hide a bit far from current experimental sensitivity. A clear example of the latter is when the WIMPS interacts with SM particle through a pseudo-scalar field \cite{Arcadi:2017kky}.

Now we have briefly introduced WIMPs, we will address some recent results and trends in the field. 

\section{A Step Toward the Nature of Dark Matter with Direct Detection Experiments Only}

%\subsection{Direct Detection}

The presence of dark matter in our galaxy has been established via its gravitational effects. Though subject to uncertainties, the local dark matter density is estimated to be $0.4\, GeV/cm^3$ \cite{Catena:2009mf}. Since dark matter particles are known to be non-relativistic today, we can use the relation $\rho = n_{DM}\, m_{DM}$, to realize $n_{DM} =m_{DM}\, 0.4\, GeV/cm^3$. In natural units we get $cm^{-1} = 3 \times 10^{10} s^{-1}$, therefore $n_{DM} \sim   10^{10}/(cm^2\, s)\,\,\, (GeV)/m_{DM})$. One may not realize at first but it means that if a human body can be thought as a $1m^2$ box, we have about $10^{14}$ WIMPs passing it through per second. This is literally mind-blowing. Although, these particles interact weakly with ordinary matter, making their probability to scatter off nuclei in our body very small, explaining why we have not caught one yet.

In the attempt to detect direct signs of WIMPs, underground detectors are built using all sort of targets (Xe, Ge, Na, I, W,  O, Ar ). They search for either electronic or nuclear scatterings at low energies. Keeping their background under control they are powerful machines in the search for WIMPS. Setting aside the controversial DAMA modulation signal, we have not observed any signal in direct detection experiments. 

\begin{figure}
\centering
\includegraphics[scale=0.5]{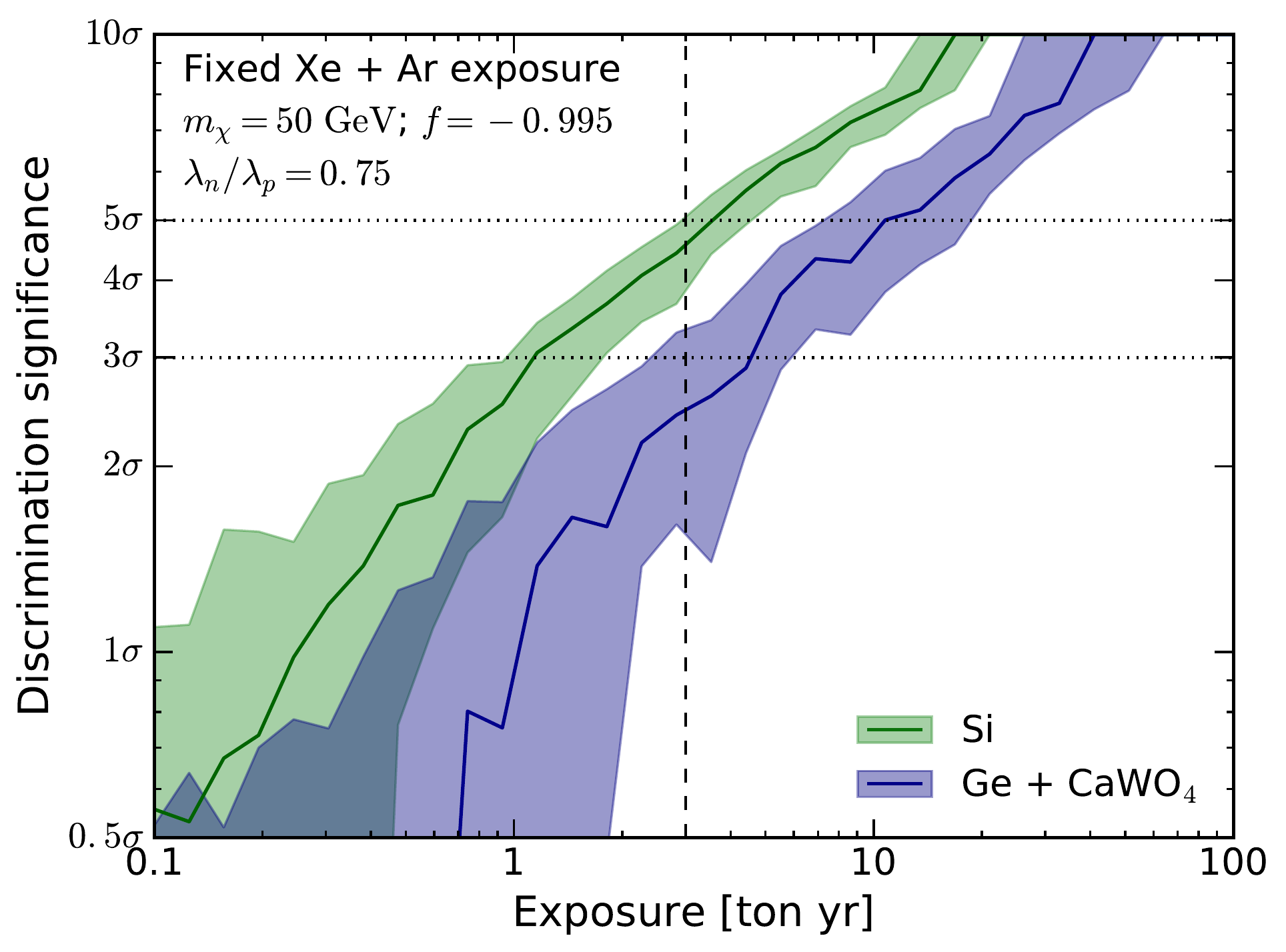}
\caption{Distinguing the particle vs antiparticle nature of WIMPs using direct detection experiments only. In the figure we assumed a $50$~GeV WIMP.}
\label{fig1}
\end{figure}

Being optimistic, if we eventually observe a signal in direct dark matter detection experiments, what will we learn? Several works have been done in this direction showing how precisely we can eventually reconstruct the WIMP-nucleon scattering cross section depending on the target and number of events observed. Is that all one can do? We are all devoted to one goal which is to unveil the nature of dark matter.  So what does it take? All we need is a bit of luck and signals at three different targets. For Majorana particles, at zero-momentum transfer limit, the WIMP-nucleus scattering cross section is \cite{Cerdeno:2010jj}, 

\begin{equation}
\sigma^M_{\rm SI}= \frac{4\mu_{\chi A}^2}{\pi}\left[\lambda_{p}^M\,N_p + \lambda_{n}^M\, N_n\right]^2 \,.\label{eq:sigmaM}
\end{equation}where $ \mu_{\chi A}=M_\chi M_A/(M_\chi+M_A)$ is the reduced mass of the DM-nucleus system, $\lambda_p,\lambda_n$ are the dark matter coupling strength with protons and neutrons, $N_p$ is the number of protons and $N_n$ is the number of neutrons. For a given dark matter mass, from this equation, we can see that if we observe two signals at two different targets, we will have two equations and two variables. Therefore, we can solve for it, and then predict how many events a third direct detection experiment will observe if Eq.\ref{eq:sigmaM} is indeed the solution for the observations.  If the dark matter is a Dirac fermion instead, this is no longer true, because the particle is not its own antiparticle. This simple logic is not quite valid in practical terms because we have many isotopes in a given experiment, statistical uncertainties involved, etc. In summary, it is not that simple. Accounting for these effects in Fig.\ref{fig1} we show which discriminating power one can achieve depending on the targets and exposure achieved. From Fig.\ref{fig1} we conclude that with sufficiently large exposure we can assess at $5\sigma$ whether a $50$~GeV WIMP is a Dirac or Majorana particle. The same argument applies for real vs complex scalar, real or complex vector dark matter. In summary, if nature indeed favor WIMPs and we observe signals in the near future we will be able to give an important step toward the fundamental nature of dark matter with direct detection experiments only \cite{Queiroz:2016sxf,Kavanagh:2017hcl}.

\section{Multiple Signals in Indirect Dark Matter Detection  Searches}

In regions of high density of dark matter, WIMPs may self-annihilate into SM particles or other exotic final states. As result, a visible signal in gamma-rays, cosmic-rays, neutrinos, and anti-matter may be produced. Indirect dark matter detection differently from direct detection and collider searches for dark matter is subject to large uncertainties. Moreover, even if signals are observed consistent with what we expected from dark matter annihilation or decay, it is very hard to exclude other astrophysical sources as potential explanations. Making the detection of dark matter to be very challenging. That said, we will briefly review some potential dark matter signals being currently discussed in the literature. 

\subsection{GeV Gamma-ray Excess}

There are two excesses in the gamma-ray band which have been attributed to dark matter annihilation. One stemming from the galactic center \cite{Hooper:2010mq,TheFermi-LAT:2015kwa} and other from the recently discovered dwarf galaxy known as Reticulum II \cite{Geringer-Sameth:2015lua}. Only the former is statistically significant. It is intriguing that both excesses favor similar dark matter properties, i.e. $\sigma v \sim 10^{-26} cm^3s^{-1}$, $M_{\chi} \sim 30$~GeV.  Current bounds from Fermi-LAT via the non-observation of gamma-ray emission in other dwarf galaxies severely restricts the dark matter interpretation of such signals \cite{Fermi-LAT:2016uux,TheFermi-LAT:2017vmf}, but unfortunately, Fermi-LAT will not conclusively address them in its lifetime of a few years. 
\subsection{GeV Antiproton Excess}

Recently an excess in antiproton has been reported using AMS-02 data \cite{Cui:2016ppb,Cuoco:2016eej}. Performing a global fit to both signal and background an excess has been observed. It again favors a dark matter annihilation cross section and mass similar to those from gamma-ray probes. These results are subject to relevant systematic uncertainties and background modelling \cite{Cholis:2017qlb,Niu:2017qfv,Yuan:2017ozr}. There is also a claim, not statistically significant, that an excess in Anti-Helium has been spotted \cite{Coogan:2017pwt}. Fortunately, AMS-02 will continue to run for a sufficiently long time that will allow us to assess whether these observations are due to dark matter.

\subsection{X-ray Excess}

The excesses aforementioned are all tied to dark matter annihilation. However, a $keV$ line emission has been observed with the XMM Newton satellite, and it constitutes a potential signal of decaying dark matter \cite{Bulbul:2014sua,Boyarsky:2014jta}. This decaying dark matter interpretation is also fiercely constrained by other dataset \cite{Jeltema:2015mee,Ruchayskiy:2015onc}, but fortunately the origin for this x-ray excess will likely be addressed in a short timescale with Micro-X \cite{Figueroa-Feliciano:2015gwa}.

\section{Prospects}

Regarding the signals discussed previously, the GeV gamma-ray excess in the galactic center is unlikely to be resolved in the near future due to the lack of experimental sensitivity, conversely to the $keV$ line. The other excesses are not as robust and need further scrutiny to see if they persist. 
On the theoretical side though, the next decade will be critical. Most models of WIMPs with masses below $1$~TeV are expected to be fully probed by the next generation of direct detection experiments. Hence, if we do observe clear signals we should arguably push for directional direct detection experiments. If we do not, WIMPs should be heavier than a few TeV and/or interact with Standard Model particles in non-trivial ways, for instance via more than one vector or scalar mediator, or through a pseudo-scalar, among others. These setups appear in a multitude of models, thus still rendering WIMPs as compelling dark matter candidates. Although, we emphasize that if direct detection experiments manage to exclude spin-independent WIMP-nucleon scattering cross section all the way down to $\sim 10^{-51} cm^2$, we will probably have to abandon the thermal production paradigm because very few WIMP models below the $1$~TeV scale will survive. For such models, WIMPs will lose predictivity and most of its theoretical appeal.

\end{document}